\documentclass{article}

\usepackage{arxiv}

\usepackage[utf8]{inputenc} 
\usepackage[T1]{fontenc}    
\usepackage{hyperref}       
\usepackage{url}            
\usepackage{booktabs}       
\usepackage{amsfonts}       
\usepackage{nicefrac}       
\usepackage{microtype}      
\usepackage{lipsum}
\usepackage{graphicx}

\usepackage[utf8]{inputenc} 
\usepackage[T1]{fontenc}    
\usepackage{hyperref}       
\usepackage{url}            
\usepackage{booktabs}       
\usepackage{amsfonts}       
\usepackage{nicefrac}       
\usepackage{microtype}      
\usepackage{xcolor}         

\usepackage{pifont}
\usepackage{graphicx}
\usepackage{subcaption}
\usepackage{amsmath}
\usepackage{multirow}
\usepackage{float}

\usepackage{enumitem}
\setlist[itemize]{nosep, itemsep=1pt, topsep=1pt, leftmargin=1em, labelwidth=*}
\setlist[enumerate]{nosep, itemsep=1pt, topsep=1pt, leftmargin=1em, labelwidth=*}

\newcommand{\mysection}[1]{\vspace{-.1in}\section{#1}\vspace{-.05in}}
\newcommand{\mysubsection}[1]{\vspace{-.06in}\subsection{#1}\vspace{-.06in}}

\title{Still Camouflage, Moving Illusion: View-Induced Trajectory Manipulation in Autonomous Driving}

\author{%
    {\bfseries
    Shuo Ju$^{1}$, 
    Qingzhao Zhang$^{2}$, 
    Huashan Chen$^{1}$, 
    Xuheng Wang$^{3}$, 
    Haotang Li$^{2}$,} \\
    {\bfseries
    Wanqian Zhang$^{1}$, 
    Feng Liu$^{1}$, 
    Kebin Peng$^{4}$, 
    and Sen He$^{2}$%
    }\\[0.3cm]
    {\normalfont $^{1}$ Institute of Information Engineering, Chinese Academy of Sciences} \\
    {\normalfont $^{2}$ The University of Arizona} \\
    {\normalfont $^{3}$ Beijing Jiaotong University} \\
    {\normalfont $^{4}$ East Carolina University}
}

\begin{document}
\maketitle

\begin{abstract}
Existing physical adversarial attacks on vision-based autonomous driving induce time-evolving perception errors, including biased object tracking or trajectory prediction, through (i) sophisticated physical patch inducing detection box drift when entering the view distance, or (ii) dynamically changing patches that cause different perception errors at different time. In both cases, viewing-angle variation is treated as a challenge, requiring adversarial patches to remain effective across frames under varying views, leading to complex multi-view optimization.

In contrast, we show that viewing-angle variation itself can be turned into an attack tool. We design a new attack paradigm where a static, passive adversarial camouflage is mounted on a vehicle whose view-dependent appearance naturally evolves with relative motion, inducing consistent feature drift across frames. This causes the system to infer a physically plausible but incorrect trajectory, such as a false cut-in, which propagates to downstream decision-making and triggers unnecessary braking. Unlike prior approaches that require multi-view robustness or active intervention, our attack emerges from normal driving dynamics and is easy to deploy: a parked vehicle with a natural camouflage can induce hard braking in passing autonomous vehicles.
We demonstrate the novel attack on nuScenes dataset, showing the effectiveness with an end-to-end success rate of up to 87.5\%, measured by hard-braking events, and robustness across different scene backgrounds, victim vehicle speeds, and perception models.
\end{abstract}

\section{Introduction}
\label{sec:intro}

Vision-based autonomous driving systems rely on temporal perception: detections from consecutive frames are associated by tracking, converted into future trajectories by prediction, and finally consumed by planning~\cite{mao20233d, omeiza2021explanations, grigorescu2020survey}. This architecture ensures the stability of the decision-making process, meaning that a single-frame perception error may not necessarily propagate through the downstream pipeline~\cite{wang2023does, Jia2020Fooling}. Therefore, for an adversarial attack to be successful, it must exhibit coherent temporal evolution to survive downstream aggregation and influence planning-level decisions.

However, existing physical attacks mainly focus on either perception-level manipulation under single-frame settings, such as inducing object hiding~\cite{huang2020upc, zhu2023tpatch} or misclassification~\cite{man2023person, Lou2025pga}, or tracking-level hijacking across multiple frames~\cite{chahe2024rss, ma2023wip, ma2025controlloc, muller2022attackzone}. Perception-oriented attacks generally optimize perturbations to remain effective under changing viewing angles and distances, while tracking-oriented attacks explicitly manipulate temporal states or tracker associations. Despite their differences, these methods share one common assumption: viewing-angle variation is treated as a robustness challenge that the attack should remain invariant against. They overlook a critical attack surface in driving scenarios: \textit{the natural viewing-angle variation during relative motion can itself become the source of temporally coherent perception manipulation.} This motivates us to explore the possibility of a ``still camouflage, moving illusion'': \textit{leveraging natural \textbf{viewing-angle variation} to transform static spatial patterns into dynamic temporal perturbations, thereby creating coherent motion illusions.}

\begin{figure}[t]
    \centering
    \begin{subfigure}{0.5\textwidth}
        \centering
        \includegraphics[width=0.9\textwidth]{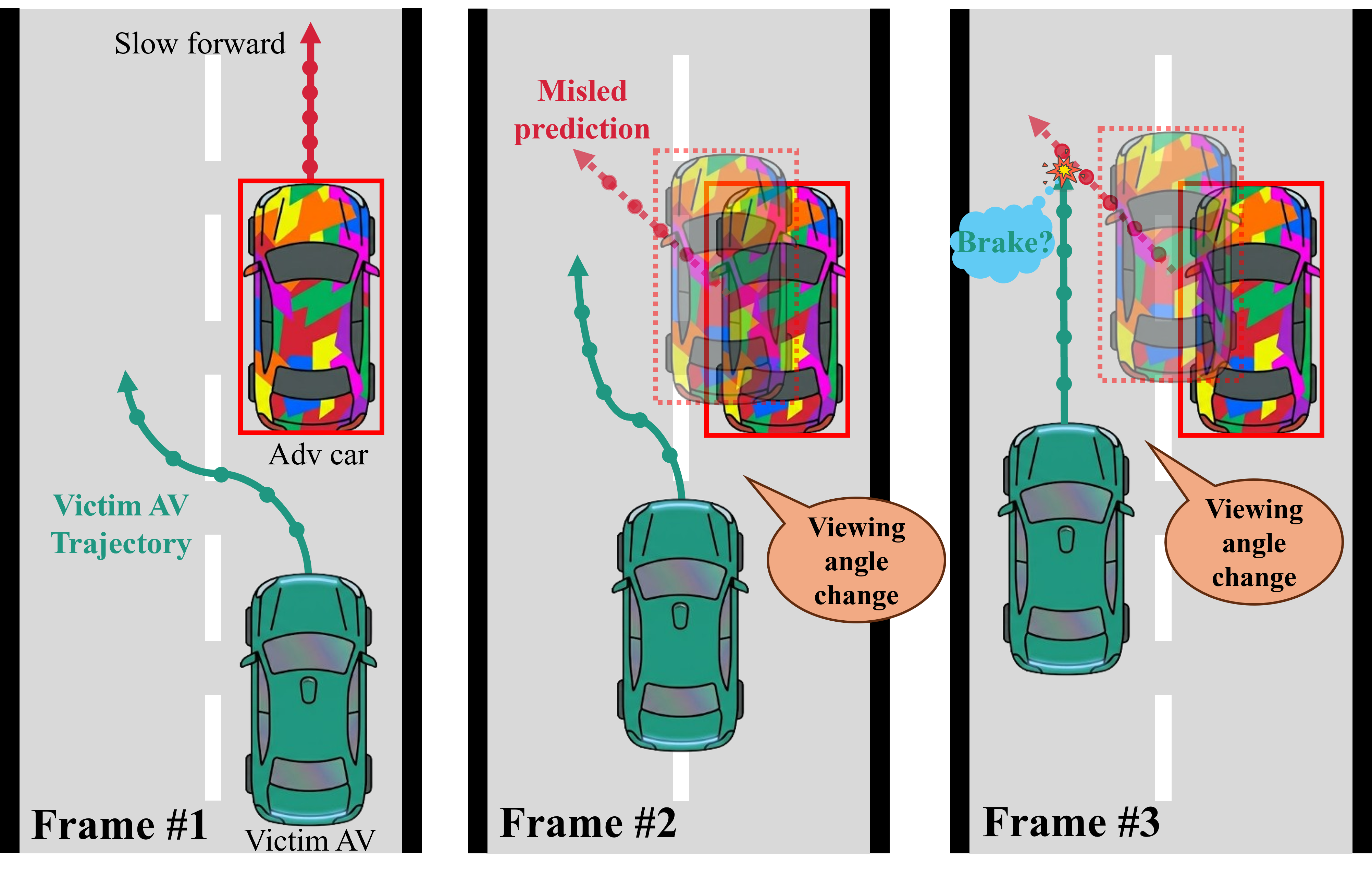}
        \caption{Blocking overtaking $\to$ gaining advantage on road.}
        \label{fig:attack-scenario-1}
    \end{subfigure}
    \hfill
    \begin{subfigure}{0.447\textwidth}
        \centering
        \includegraphics[width=0.9\textwidth]{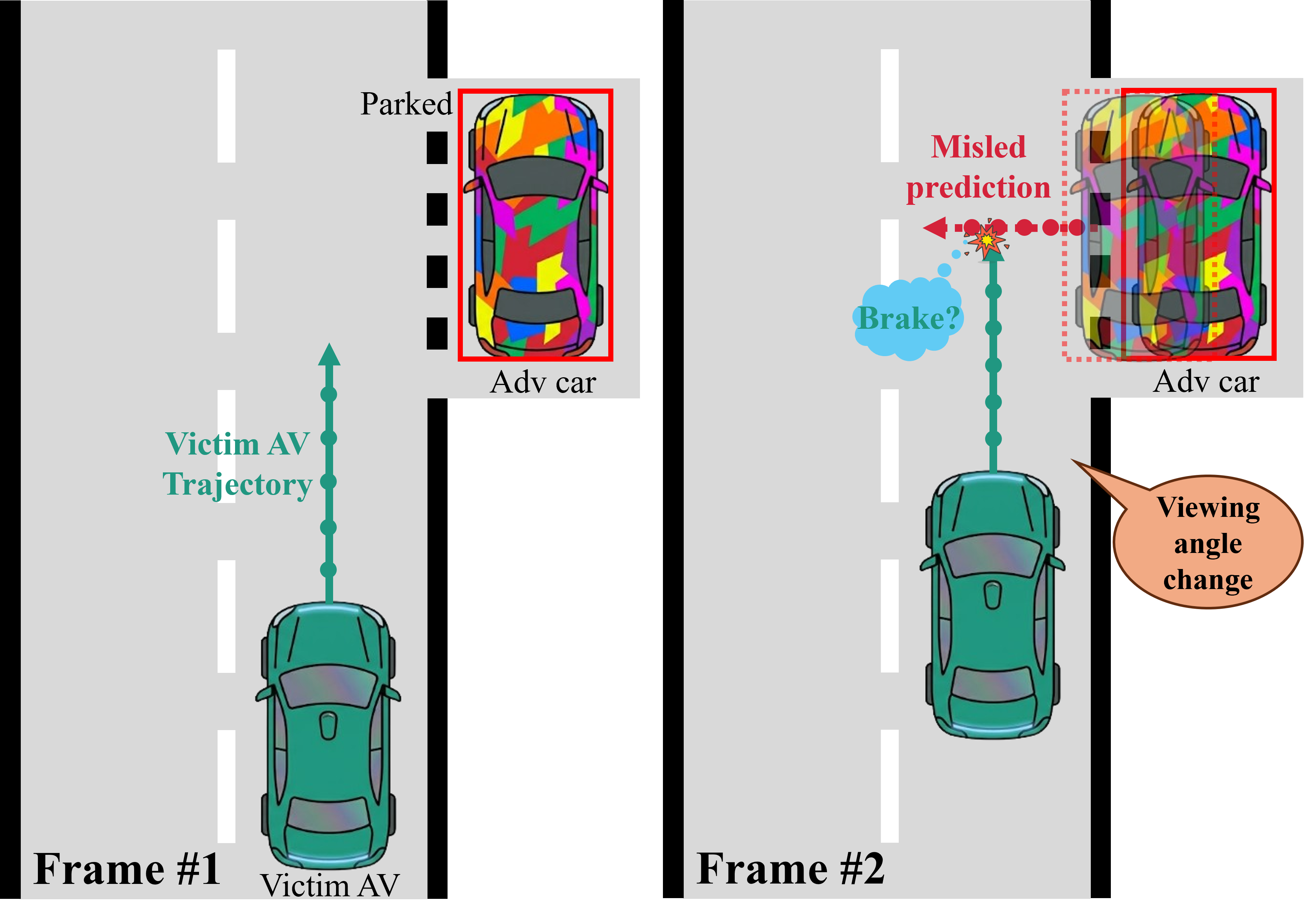}
        \caption{Causing hard braking $\to$ safety hazards.}
        \label{fig:attack-scenario-2}
    \end{subfigure}
    \caption{Attack demonstration: static camouflage on an adversarial vehicle induces temporally coherent perception errors that propagate through tracking and prediction, ultimately triggering harmful driving decisions in the victim autonomous vehicle.}
    \label{fig:attack-scenario}
\vspace{-.15in}
\end{figure}

Our attack is motivated by a simple physical observation, as illustrated in Fig.~\ref{fig:attack-scenario}. When an autonomous vehicle (AV) moves relative to a target vehicle, the target is naturally observed from changing viewing angles. As a result, the projected appearance of a static adversarial camouflage evolves across frames, even though the physical camouflage itself remains unchanged. Rather than suppressing this view-dependent variation, we exploit it to induce a sequence of consistent detection shifts. Such shifts can be interpreted by downstream tracking and prediction as a plausible but false motion pattern, e.g., a vehicle gradually cutting into the ego path. Since the camouflage is physically static, the attack requires no dynamic display, active projection, or frame-wise intervention.
Importantly, this motion illusion can force the victim AV to \textit{abandon overtaking maneuvers}, thereby giving the adversarial vehicle driving advantage, or trigger \textit{unnecessary hard braking} that creates safety-critical hazards.

However, realizing this idea is challenging. The goal is not to maximize detector errors independently in each frame, but to make a single shared physical camouflage induce temporally compatible 3D bbox shifts under changing viewing angles. This camouflage must remain physically projectable onto the vehicle surface, preserve the target as a valid detected object, and make the displacement accumulate over time rather than appear as unstable frame-wise noise. To address these challenges, we optimize a static adversarial camouflage under a geometry-constrained multi-frame formulation. We first select attackable scenarios and frame sequences with sufficient viewing-angle variation, and render the shared camouflage through a physically consistent mesh and UV pipeline. We then identify the optimal displacement direction and optimize the shared camouflage with motion-aware losses to drive progressive 3D bbox displacement under real projection constraints.

Experiments on nuScenes~\cite{nuscenes} show that our attack achieves an attack success rate (ASR) of up to 87.5\%, demonstrating that normal viewing-angle variation during relative motion can be exploited to create coherent moving illusions from a static physical camouflage. We show that the induced motion illusion leads to realistic driving impacts such as abandoned overtaking and unnecessary hard braking. Further analysis shows that the attack transfers effectively across vehicle types, becomes stronger under larger viewing-angle variation, larger projection area, brighter illumination, closer distance, and lower relative speed, and consistently benefits from all proposed modules and optimization objectives.

    
    

\textbf{Our work makes two contributions:} (1) We identify viewing-angle variation as a new attack mechanism for autonomous driving systems and propose a geometry-constrained framework for generating static camouflage that induces progressive 3D bounding-box movement across frames. (2) Extensive evaluation on nuScenes shows that the induced motion illusion propagates to downstream planning and triggers realistic impacts such as abandoned overtaking and unnecessary hard braking.
\section{Related Work}

\begin{table}[t]
\caption{Comparison between prior related works and our method.}
\label{tab:related-work}

\resizebox{\textwidth}{!}{%
\begin{tabular}{ccccccc}
\hline
\textbf{Related Work} & \textbf{\begin{tabular}[c]{@{}c@{}}Viewing-Angle\\ Dependency\end{tabular}} & \textbf{\begin{tabular}[c]{@{}c@{}}Spatio-temporal\\ Consistency\end{tabular}} & \textbf{\begin{tabular}[c]{@{}c@{}}2D/3D BBox\\ Manipulation\end{tabular}} & \textbf{\begin{tabular}[c]{@{}c@{}}Attack\\ Pattern\end{tabular}} & \textbf{\begin{tabular}[c]{@{}c@{}}Pattern\\ Variability\end{tabular}} & \textbf{\begin{tabular}[c]{@{}c@{}}System-level\\ Impact\end{tabular}} \\ \hline
Jia et al.\cite{Jia2020Fooling} & \ding{55} & \ding{51} & 2D \& move & patch & static & \ding{55} \\
AttrackZone\cite{muller2022attackzone} & \ding{55} & \ding{51} & 2D \& move & patch & dynamic & \ding{55} \\
FCA\cite{wang2022fca} & \ding{55} & \ding{55} & 3D \& hiding & camouflage & static & \ding{55} \\
DTA\cite{suryanto2022dta} & \ding{55} & \ding{55} & 3D \& hiding & camouflage & static & \ding{55} \\
ACTIVE\cite{suryanto2023active} & \ding{55} & \ding{55} & 3D \& hiding & camouflage & static & \ding{55} \\
ControlLoc\cite{ma2025controlloc} & \ding{55} & \ding{51} & 2D \& move & patch & dynamic & \ding{51} \\
Lou et al.\cite{Lou2025pga} & \ding{55} & \ding{51} & 3D \& move & physical object & static & \ding{51} \\
\textbf{OURS} & \textbf{\ding{51}} & \textbf{\ding{51}} & \textbf{3D \& move} & \textbf{camouflage} & \textbf{static} & \textbf{\ding{51}} \\ \hline
\end{tabular}%
}
\vspace{-.15in}
\end{table}

\textbf{Autonomous driving systems.}
In recent years, autonomous driving (AD) has advanced substantially, leading to systems such as Tesla Full Self-Driving~\cite{tesla2026} and Baidu Apollo~\cite{apollo2022}. Traditional AD systems typically adopt a cascaded pipeline of perception, tracking, prediction, and planning, while end-to-end systems such as ST-P3~\cite{hu2022stp3}, UniAD~\cite{hu2023uniad}, and VAD~\cite{jiang2023vad} learn driving-oriented representations directly from sensor observations. Recent Vision-Language-Action (VLA) systems, such as Alpamayo~\cite{alpamayo2026}, further integrate visual perception, reasoning, and action generation into a unified architecture. Despite these architectural differences, AD systems still rely on temporally consistent perception of surrounding agents, making persistent perception bias safety-critical.

\textbf{3D camouflage attacks.}
Compared with localized patches, 3D camouflage better conforms to vehicle geometry and offers improved physical realism and stealth~\cite{zhang2018camouflage, huang2020upc, wang2021dual}. Existing camouflage attacks against AD perception~\cite{suryanto2022dta, wang2022fca, suryanto2023active, Lou2025pga}, which mainly target perception-level errors such as object hiding or misclassification, treat viewing-angle variation as a challenge. In contrast, our work directly exploits viewing-angle variation as the attack mechanism, where the same static camouflage produces view-dependent appearance evolution across consecutive frames, thereby inducing progressive 3D bounding-box movement.

\textbf{Tracker hijacking attacks.}
Prior studies show that attacking isolated detections alone may be insufficient because tracking systems aggregate temporal observations and can suppress unstable single-frame errors~\cite{wang2023does}. Jia et al.~\cite{Jia2020Fooling} first studied adversarial attacks against multiple object tracking (MOT) and proposed tracker hijacking by manipulating detection boxes to steer tracker trajectories. AttrackZone~\cite{muller2022attackzone} moved tracker hijacking into the physical world by projecting adversarial perturbations to hijack Siamese trackers, while ControlLoc~\cite{ma2025controlloc} optimized physical perturbations to alter perceived object location and shape in camera-based AD systems. These works highlight temporal perception manipulation beyond isolated detector outputs, but mainly target 2D tracking bbox and rely on dynamic deployment. In contrast, our attack uses static camouflage and natural viewing-angle variation to induce temporally coherent 3D bbox displacement, which further propagates to planning, making it easier to deploy and closer to realistic driving scenarios.

\textbf{Prediction-oriented physical attacks.}
Recent works further investigate how perception errors propagate to downstream prediction and planning~\cite{zhang2022adversarial}. Lou et al.~\cite{lou2024first} perturb LiDAR perception by placing physical objects near parked vehicles, causing bbox errors that propagate through tracking and prediction. Yu et al.~\cite{yu2025enduring} further improve attack persistence and robustness with an optimization-based perturbation framework. However, these works mainly rely on LiDAR-oriented perturbations or external object placement. In contrast, our attack attaches a single static camouflage directly onto the adversarial vehicle and exploits natural viewing-angle variation to induce motion illusion in continuous driving scenarios.

As summarized in Tab.~\ref{tab:related-work}, our method differs from prior works in three aspects: it uses a fully static camouflage for passive deployment, exploits viewing-angle variation as the temporal attack mechanism, and induces spatio-temporally consistent 3D bbox movement that propagates to downstream driving behaviors.

\section{Threat Model}
\label{sec:threat-model}

\textbf{System assumptions.}
We consider a modular autonomous driving (AD) stack that uses camera-based visual inputs to support driving decisions. The system contains multiple downstream components, including sensing, perception, tracking, trajectory prediction, and motion planning. Our attack targets the perception stage by manipulating the detected 3D bounding box of a vehicle, while the downstream system naturally propagates the induced perception bias to later modules. Tesla~\cite{tesla2026} is one representative example of such a vision-based AD system.

\textbf{Adversary model.}
We assume that the attacker can attach a \textbf{static physical camouflage} to a vehicle under the attacker's control. The attack goal is to induce unsafe driving behaviors or gain driving advantage against a victim AD vehicle, depending on the target scenario discussed later. During attack optimization, the attacker has white-box access to the victim perception model, or alternatively performs transfer attacks by optimizing on a surrogate perception model and transferring the attack to an unknown target stack. We do not assume knowledge of downstream modules such as tracking, prediction, or planning. The attacker does not compromise the victim AD system through sensor spoofing, signal injection, software exploitation, or other direct system intrusion~\cite{man2020ghostimage, sp1, sp2, phantom}. Instead, the attack is passively triggered through normal relative motion between vehicles, where changing viewing angles naturally induce appearance evolution of the static camouflage.

\textbf{Attack scenarios.}
We consider two representative scenarios as illustrated in Fig.~\ref{fig:attack-scenario} and discussed in Sec.~\ref{sec:intro}: \textit{abandoned overtaking} and \textit{unnecessary hard braking}. Our attack methodology is \textbf{general} across both attack scenarios, which induces a false cut-in maneuver of the adversarial vehicle and makes the victim AD vehicle brake or decelerate. 
\section{Design}
\mysubsection{Problem Formulation}
\label{sec:problem}

We formulate the proposed attack as a static camouflage attack problem over sequential multi-view observations. Let $\mathcal{D}$ denote the original dataset. The synchronized multi-view observation at timestamp $t$ is $\mathcal{X}_{t}=\{I_t^{(v)}\}_{v=1}^{V}$, where $V$ denotes the number of camera views and is set to 6. Let $\mathcal{O}=\{\mathcal{X}_{t}\}_{t=1}^{T}$ denote an original sequence selected from $\mathcal{D}$. Then we select a valid consecutive $K$-frame sequence $\mathcal{S}=\{\mathcal{X}_{k}\}_{k=1}^{K}$ from $\mathcal{O}$. Based on prior studies showing successful attacks on only a few consecutive frames~\cite{Jia2020Fooling,muller2022attackzone,ma2025controlloc,lou2024first}, we set $K=3$, which is sufficient to propagate the perception bias to the downstream pipeline. 

Let $\mathcal{V}$ denote the target vehicle and $\mathcal{G}$ its vehicle mesh. The attack optimizes a static physical camouflage $\delta$, parameterized on the renderable region of $\mathcal{G}$, and generates adversarial observations $\tilde{\mathcal{X}}_{k}$ over $\mathcal{S}$. Since $\delta$ is static, its appearance variation across frames is induced by viewing-angle variation under relative motion.
Let $D(\cdot)$ denote the target 3D detector and $\Pi(\cdot)$ select the detected 3D bbox of $\mathcal{V}$. The attacked box at frame $k$ is $\hat{\mathbf{b}}_{k}=\Pi(D(\tilde{\mathcal{X}}_{k}))$, while the corresponding clean box is $\mathbf{b}_{k}$. Their centers are denoted by $\hat{\mathbf{c}}_{k}$ and $\mathbf{c}_{k}$, respectively. The attack aims to move  $\hat{\mathbf{c}}_{k}$ in a temporally coherent manner with progressively larger displacements across $\mathcal{S}$ while keeping $\hat{\mathbf{b}}_{k}$ detectable.

\mysubsection{Attack Design Overview}
\label{sec:overview}

\begin{figure}[t]
    \centering
    \includegraphics[width=0.9\textwidth]{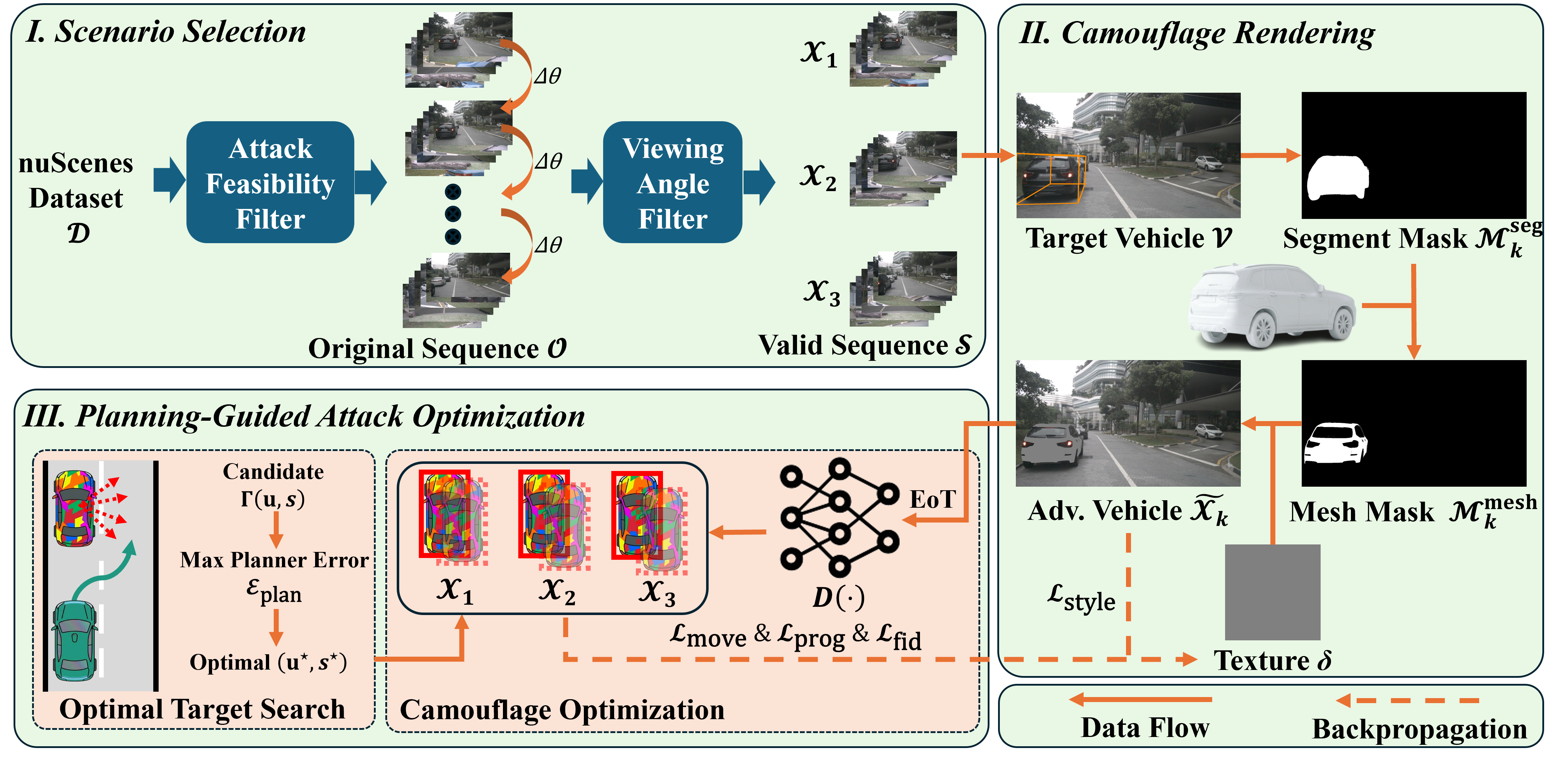}
    \caption{Pipeline of the proposed attack. We first select a valid consecutive 3-frame sequence $\mathcal{S}$, render a static camouflage $\delta$ onto the renderable region of $\mathcal{V}$, and optimize $\delta$ under planning-guided objectives to induce progressive 3D bbox displacement.}
    \label{fig:pipeline}
\vspace{-.15in}
\end{figure}

Inspired by viewing-angle variation naturally induced by relative motion in driving scenarios, we design a static physical camouflage attack that exploits this variation to produce temporally coherent 3D detection bias. As illustrated in Fig.~\ref{fig:pipeline}, the design follows three stages. \textit{Scenario Selection} identifies sequences where the target vehicle is attack-relevant and exhibits sufficient viewing-angle variation. \textit{Camouflage Rendering} maps the static camouflage $\delta$ to the renderable region of $\mathcal{G}$ while preserving visible-surface constraints. \textit{Planning-Guided Attack Optimization} optimizes $\delta$ so that the biased detections form progressive and coherent 3D bbox movements that can propagate to the downstream pipeline.

\mysubsection{Scenario Selection and Camouflage Rendering}
\label{sec:selection}

To ensure the effectiveness of the attack, we employ a two-stage filtering process to select the valid consecutive sequence $\mathcal{S}$ from the dataset $\mathcal{D}$.

\textbf{Attack Feasibility Filter (AFF).}
We first select an original sequence $\mathcal{O}$ in which the target vehicle $\mathcal{V}$ is continuously visible, remains in front of the ego vehicle, and is relevant to the ego future path without already intruding into it in the clean scene. This ensures the attack has the potential to trigger hazardous planning responses by introducing detection displacement into a previously safe but highly relevant region of the ego-vehicle trajectory.

\textbf{Viewing-Angle Filter (VAF).}
The core of our mechanism relies on viewing-angle variation. Accordingly, VAF further extracts a valid consecutive $K$-frame sequence $\mathcal{S}$ from $\mathcal{O}$ by ensuring sufficient viewing-angle variation between $\mathcal{X}_1$ and $\mathcal{X}_K$. Unlike existing attacks that treat such variation as a robustness obstacle to overcome, we prefer sequences with larger viewing-angle variation, ensuring that the same static camouflage $\delta$ can produce view-dependent responses and form coherent motion illusions.

After obtaining $\mathcal{S}$, we render the static camouflage $\delta$ onto $\mathcal{V}$ using PyTorch3D~\cite{ravi2020pytorch3d} with camera intrinsics and annotations. We use $\mathcal{M}_{k}^{\mathrm{mesh}}$ to determine the precise renderable region and $\mathcal{M}_{k}^{\mathrm{seg}}$, generated by SAM2~\cite{sam2025}, to handle occlusions.

\mysubsection{Planning-Guided Attack Optimization}
\label{sec:optimization}

This stage optimizes the camouflage $\delta$ by first searching for a planning-guided displacement target and then realizing it through detector-level optimization. The downstream pipeline is used to determine a hazardous target direction and step size, while the optimization itself is applied to the target 3D detector.

\textbf{Optimal Target Search.}
To specify the target displacement pattern, we perform a planning-guided target search over a candidate direction set $\mathcal{U}$ and a candidate step-size set $\mathcal{D}_{s}$. Let $\Gamma(\mathbf{u},s)$ be the fake prediction generated by direction $\mathbf{u}$ and step size $s$, and let $\mathcal{E}_{\mathrm{plan}}(\cdot)$ measure the induced planning error. The optimal direction and step size are defined as
\vspace{-5pt}
\begin{equation}
    \label{equ:optimal-direction}
    (\mathbf{u}^{\star}, s^{\star})
    =
    \arg\max_{\mathbf{u}\in\mathcal{U},\,s\in\mathcal{D}_{s}}
    \mathcal{E}_{\mathrm{plan}}
    \left(
    \Gamma(\mathbf{u},s)
    \right).
\end{equation}

Since $\delta$ is shared within a training group, we average the searched targets and obtain the group-level attack target $(\bar{\mathbf{u}},\bar{s})$. The 3D bbox displacement at frame $k$ along the target direction is defined as
\vspace{-5pt}
\begin{equation}
    d_k
    =
    \left(
    \hat{\mathbf{c}}_{k}-\mathbf{c}_{k}
    \right)^{\top}
    \bar{\mathbf{u}},
    \qquad k=1,\ldots,K .
\end{equation}

The planning-guided optimization target is to maximize the overall target-direction displacement while enforcing step consistency across consecutive frames:
\vspace{-5pt}
\begin{equation}
    \max_{\delta}
    \sum_{k=1}^{K} d_k
    -
    \lambda
    \sum_{k=1}^{K-1}
    \left(
    d_{k+1}-d_k-\bar{s}
    \right)^2 .
\end{equation}

\textbf{Optimization Objectives.}
We implement the planning-guided target using move and progress losses, and regularize the optimization with fidelity and style losses. To improve physical robustness, we further apply 3D Expectation over Transformation (EoT)~\cite{athalye2018eot} during optimization by introducing random perturbations to the target vehicle pose and projection, including yaw perturbation, 3D translation, depth-ratio variation, and 3D scale variation. The losses are defined below.

\begin{enumerate}[leftmargin=*]

\item \textit{Move loss.}
The move loss drives the target 3D bbox center along the planning-guided direction $\bar{\mathbf{u}}$, i.e., maximizing the target-direction displacement $d_k$ defined above.

\item \textit{Progress loss.}
The progress loss incorporates the planning-guided frame-wise step size $\bar{s}$ and encourages a progressively increasing, spatio-temporally consistent 3D bbox movement. We constrain the displacement increment between consecutive frames to match $\bar{s}$.

\item \textit{Fidelity loss.}
The fidelity loss preserves detection credibility by penalizing deviations in confidence, length, width, height, and yaw, preventing target-vehicle disappearance or severe box distortion. Let $\mathbf{r}$ and $\hat{\mathbf{r}}$ denote the clean and attacked fidelity attribute vectors, respectively.

\item \textit{Style loss.}
The style loss $\mathcal{L}_{\mathrm{style}}$ combines total variation loss $\mathcal{L}_{\mathrm{tv}}$, which reduces high-frequency noise, and non-printability-score loss $\mathcal{L}_{\mathrm{nps}}$, which constrains optimized colors to printable ranges.

\end{enumerate}

\vspace{-.2in}
\begin{equation}
\begin{aligned}
\mathcal{L}_{\mathrm{move}}
&= - \sum_{k=1}^{K} d_k ,
\qquad
\mathcal{L}_{\mathrm{prog}}
= \sum_{k=1}^{K-1}
\left( d_{k+1} - d_k - \bar{s} \right)^2 ,
\qquad
\mathcal{L}_{\mathrm{fid}}
= \left\| \hat{\mathbf{r}} - \mathbf{r} \right\| .
\end{aligned}
\end{equation}

\section{Evaluation}
\mysubsection{Evaluation Setups}
\label{sec:exp_setup}

\textbf{Datasets.} 
We evaluate our attack on nuScenes~\cite{nuscenes} using 220 attack scenarios, each with a contiguous sequence of camera feeds and annotations, spanning different combinations of target vehicle type, relative position, and driving direction. Specifically, the target vehicle may be an SUV, sedan, or van, located in either the front-left (\textit{L}) or front-right (\textit{R}) region of the ego vehicle, and traveling in either the same (\textit{S}) or opposite (\textit{O}) direction. We denote each category using the format \textit{TYPE-POS-DIR}; for example, \textit{SUV-L-S} represents an SUV located in the front-left region of the ego vehicle and driving in the same direction. Such categories naturally involve attack scenarios in Fig.~\ref{fig:attack-scenario}.

\textbf{Evaluation settings.}
We evaluate the attack under two settings: \textit{cross-validation} and \textit{specific-scenario} attack. For the two primary categories, \textit{SUV-R-S} and \textit{SEDAN-R-S}, we perform 5-fold cross-validation, where each fold optimizes one shared static camouflage on 4/5 of the scenarios and evaluates it on the remaining 1/5. This setting measures whether the learned camouflage generalizes to unseen scenarios within the same category. For the remaining categories, due to the limited number of scenarios, we optimize and evaluate the camouflage on the same scenario, corresponding to a scenario-specific attack where the attacker targets a chosen driving situation.

\textbf{AD Models.} 
We employ BEVDet~\cite{huang2021bevdet}, BEVDepth~\cite{li2023bevdepth}, and FastBEV++~\cite{fastbev} as the vision-based 3D detectors. To evaluate attack propagation, we construct two downstream pipelines: Pipeline A comprises AB3DMOT~\cite{weng2020ab3d}, Trajectron++~\cite{2020trajectron}, and an MPC-based planner~\cite{mpc}; Pipeline B integrates CenterTrack~\cite{zhou2020ct}, HiVT~\cite{zhou2022hivt}, and an FOT-based planner~\cite{fot}.

\textbf{Metrics.}
We evaluate the attack at the perception, prediction, and planning levels using the metrics:
\begin{itemize}
    \item \textit{Frame-wise bbox displacement ($d_1,d_2,d_3$)}
    denotes the 3D bounding-box center displacement at the three attack frames, measured in meters along the split-specific average optimal direction $\bar{\mathbf{u}}$ in Eq.~\eqref{equ:optimal-direction}, which is estimated from each training split and fixed for its corresponding test split.

    \item \textit{Progressive Displacement Rate (PDR)}
    measures the percentage of samples satisfying $d_1 < d_2 < d_3$, indicating temporally consistent bbox displacement across attack frames.

    \item \textit{Average Prediction Error (APE)}
    measures the average deviation between the clean and attacked predicted trajectories of the target vehicle.

    \item \textit{Minimum Trajectory Distance (MTD)}
    measures the minimum distance between the attacked predicted trajectory of the target vehicle and the victim AV's original planned trajectory. Smaller MTD indicates stronger planning interference.

    \item \textit{Maximum Braking Deceleration (MBD)}
    measures the maximum braking deceleration triggered by the victim AV during the three attack frames.

    \item \textit{Attack Success Rate (ASR)} is the percentage of successful attacks.
    An attack is considered successful if the victim AV triggers hard braking within the three attack frames, defined as $\mathrm{MBD} \geq 3.0\,\mathrm{m/s^2}$.
\end{itemize}

\textbf{Implementation Details.} 
We optimize the static camouflage using Adam~\cite{adam} with a learning rate of 0.01 and a texture resolution of $1024 \times 1024$. For each optimized camouflage, we run 3000 steps in the cross-validation setting and 500 steps in the specific-scenario setting. The attachable region covers the vehicle body surface while excluding tires, windows, and mirrors. All experiments are conducted on a single NVIDIA H100 GPU.

\mysubsection{Attack Effectiveness}
\label{sec:main-results}

Tab.~\ref{tab:main-results} reports the main attack results on the two primary cases under 5-fold cross-validation. 
The proposed attack achieves up to 0.70 m progressive 3D bbox displacement across three detectors and triggers hard braking with up to 66.7\% ASR through two downstream pipelines.

\textbf{The attack induces temporally coherent perception bias.}
The results confirm that a static camouflage can reliably manipulate the target 3D bbox without dynamic patterns or online intervention. The average PDR reaches $70.8\%$, indicating that the induced bbox displacement is often temporally progressive rather than a single-frame perturbation. This supports our design goal of generating coherent motion bias from frame-to-frame viewing-angle variation.

\textbf{The perception bias propagates to downstream prediction and planning.}
Under Pipeline~A/B, the attack produces average APEs of $2.84/3.02$~m, showing substantial deviation from the clean target-vehicle prediction. The average MTDs of $1.67/1.48$~m further indicate that the attacked prediction approaches the victim AV's original planned path, creating a safety-critical planning conflict. As a result, the attack achieves average hard-braking ASRs of $55.3\%/62.2\%$ under Pipeline~A/B.


\begin{table}[t]
\centering
\caption{Main attack results on the two primary cases. 
We report 3D bounding-box center displacement and downstream propagation results, where values separated by `/’ correspond to Pipeline~A~/~B.}
\label{tab:main-results}
\renewcommand{\arraystretch}{1.0}
\resizebox{\textwidth}{!}{
\begin{tabular}{cccccccccc}
\toprule
\multirow{2}{*}{Scenario} & \multirow{2}{*}{Detector} & \multicolumn{4}{c}{3D BBox Displacement} & \multicolumn{4}{c}{Downstream Propagation} \\
\cmidrule(lr){3-6} \cmidrule(lr){7-10}
 & & $d_1$ (m) & $d_2$ (m) & $d_3$ (m) & PDR (\%) & APE (m) & MTD (m) & MBD (m/s$^2$) & ASR (\%) \\
\midrule
\multirow{3}{*}{SUV-R-S}   
& BEVDet    & \textbf{0.18} & \textbf{0.37} & \textbf{0.70} & \textbf{78.3} & 3.04/2.70 & 1.34/1.30 & 3.61/4.36 & 50.0/63.3 \\
& BEVDepth  & 0.13 & 0.27 & 0.46 & 66.7 & 2.08/2.53 & 2.03/1.75 & 3.19/4.20 & 53.3/61.7 \\
& FastBEV++ & 0.11 & 0.30 & 0.45 & 70.0 & 3.19/3.20 & 2.20/2.12 & 3.20/3.37 & 53.3/56.7 \\
\midrule
\multirow{3}{*}{SEDAN-R-S} 
& BEVDet    & 0.14 & 0.33 & 0.65 & 75.0 & \textbf{3.74}/\textbf{3.89} & \textbf{0.87}/\textbf{0.81} & \textbf{4.23}/\textbf{5.78} & \textbf{65.0}/\textbf{66.7} \\
& BEVDepth  & 0.07 & 0.17 & 0.39 & 61.7 & 1.94/2.65 & 2.06/1.60 & 2.95/4.23 & 48.3/61.7 \\
& FastBEV++ & 0.10 & 0.28 & 0.48 & 73.0 & 3.05/3.17 & 1.50/1.31 & 3.40/4.76 & 61.7/63.3 \\
\midrule
\multicolumn{2}{c}{Average} & 0.12 & 0.29 & 0.52 & 70.8 & 2.84/3.02 & 1.67/1.48 & 3.43/4.45 & 55.3/62.2 \\
\bottomrule
\end{tabular}
}
\vspace{-.15in}
\end{table}

\begin{figure}[t]
  \centering
  \begin{minipage}[c]{0.47\textwidth}
    \centering
    \captionof{table}{Specific-attack results on the cases with limited available scenarios.} 
    \label{tab:scenario-wise}
    \renewcommand{\arraystretch}{1.0}
    \resizebox{\textwidth}{!}{
      \begin{tabular}{lcccc}
\toprule
Scenario & Num. & $d_3$ (m) & PDR (\%) & ASR (\%) \\
\midrule
SEDAN-L-O & 19 & 0.85 & 73.7  & 68.4 \\
SEDAN-L-S & 13 & 0.79 & 84.6  & 76.9 \\
SEDAN-R-O & 14 & 0.83 & 85.7  & 71.4 \\
SUV-L-O   & 8  & 0.84 & 87.5  & 75.0 \\
SUV-L-S   & 9  & \textbf{0.96} & 77.8  & 66.7 \\
SUV-R-O   & 6  & 0.91 & 83.3  & 66.7 \\
VAN-L-O   & 4  & 0.91 & \textbf{100.0} & 75.0 \\
VAN-L-S   & 13 & 0.94 & 92.3  & 84.6 \\
VAN-R-O   & 6  & 0.93 & 83.3  & 83.3 \\
VAN-R-S   & 8  & 0.93 & 87.5  & \textbf{87.5} \\
\bottomrule
\end{tabular}
    }
  \end{minipage}
  \hfill 
  \begin{minipage}[c]{0.47\textwidth}
    \centering
    \includegraphics[width=\textwidth]{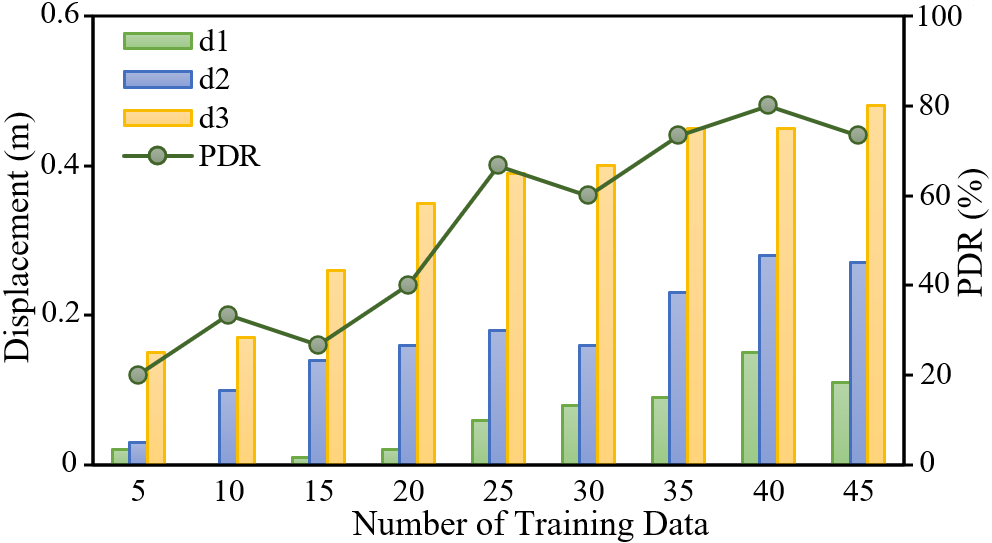}
    \caption{Effect of the number of training scenarios on cross-scenario generalization.}
    \label{fig:train-num}
  \end{minipage}
  \vspace{-.15in}
\end{figure}

\textbf{Successful attacks require sufficient cross-scenario training diversity.}
We further study the training-data requirement for achieving successful attacks on the two largest cases. For each case, we fix 15 scenarios as the test set and vary the number of training scenarios from 5 to 45. As shown in Fig.~\ref{fig:train-num}, using too few training scenarios leads to unstable test performance, indicating underfitting of the still camouflage. When the number of training scenarios reaches around 35-40, the final-frame displacement and PDR become more stable, suggesting that cross-scenario generalization requires sufficient training samples.

\textbf{The attack generalizes across diverse traffic configurations.}
For the remaining cases, the number of available scenarios is limited and unevenly distributed, making them unsuitable for train/test generalization evaluation. We therefore use a case-wise specific-attack setting, where one shared camouflage is optimized and evaluated within each case. As shown in Tab.~\ref{tab:scenario-wise}, the attack remains feasible across diverse target-vehicle types, relative positions and directions, achieving an average final-frame displacement $d_3$ of $0.89$~m, a PDR of $85.6\%$, and an ASR of $75.6\%$.

\subsection{Impact of Scenario Factors}
\label{sec:factors}

\begin{figure}[t]
    \centering
        \centering
        \includegraphics[width=\linewidth]{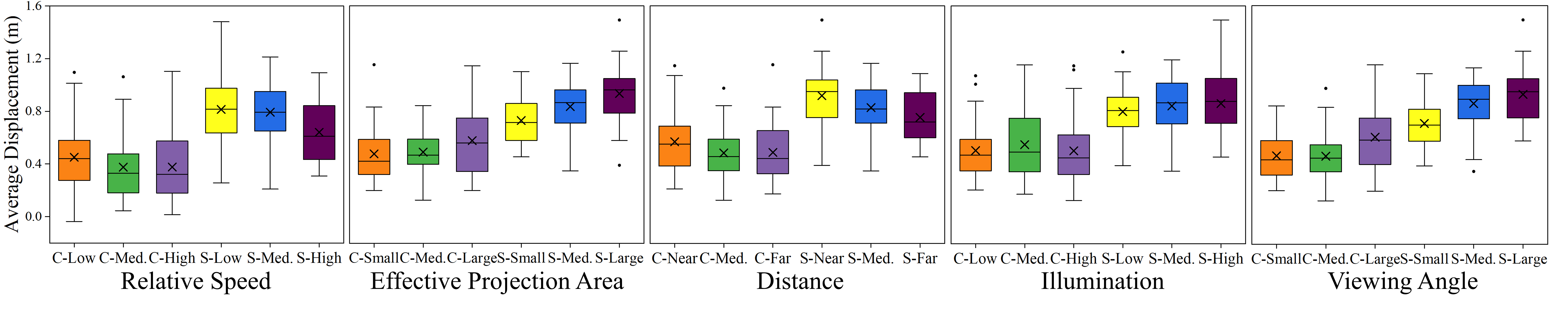}


    \vspace{-10pt}
    \caption{Impact of scenario factors---distribution of the three-frame average displacement $\bar{d}$ across different factor groups. C/S: evaluation settings cross-validation/scenario-specific.}
    \label{fig:brightness_factor}
    \vspace{-.15in}
\end{figure}

\textbf{Viewing-angle variation dominates under cross-validation.}
As shown in Fig.~\ref{fig:brightness_factor}, we analyze five scenario factors: relative speed, effective projection area, distance, illumination and viewing angle. Since real-world driving data does not provide strictly controlled variables and different factors may be correlated, we use boxplots to compare the distribution of the three-frame average displacement $\bar{d}=(d_1+d_2+d_3)/3$ under different factor groups. Under the cross-validation setting, factor effects are partially weakened by scene variation and the generalization constraint of a shared still camouflage. Nevertheless, viewing-angle variation still shows a clear distributional difference: larger viewing-angle variation generally leads to stronger displacement, which is consistent with the core mechanism of our attack.

\textbf{Scenario-specific optimization strengthens factor effects.}
Under the specific-attack setting, each scenario is optimized individually, making the factor-dependent distributions more pronounced. The attack is stronger when the target vehicle is closer, the illumination is brighter, the relative speed is lower, the effective projection area is larger, and the viewing-angle variation is larger. These observations are consistent with our design intuition: larger visible projection and closer distance increase the visual influence of the camouflage, brighter illumination improves texture visibility, lower relative speed produces smoother frame-to-frame changes, and larger viewing-angle variation amplifies the view-dependent appearance evolution of the still camouflage.

\mysubsection{Transferability}
\label{sec:transferability}

\begin{table*}[t]
\centering
\begin{tabular}{cc}
\begin{minipage}[t]{0.58\textwidth}
    \centering
\centering
\small
\setlength{\tabcolsep}{1pt}
\caption{Cross-model transferability; reported 3D bbox displacement $d_1/d_2/d_3$ (m) in three frames.}
\label{tab:transfer-model}
\renewcommand{\arraystretch}{1.0}
\begin{tabular}{lccc}
\toprule
\multirow{2}{*}{\shortstack{Source\\Detector}} 
& \multicolumn{3}{c}{Target Detector} \\
\cmidrule(lr){2-4}
& BEVDet & BEVDepth & FastBEV++ \\
\midrule
BEVDet 
& \textbf{0.16/0.35/0.68} 
& -0.09/-0.08/0.01 
& -0.03/0.06/0.05 \\
BEVDepth 
& -0.06/-0.02/-0.09 
& \textbf{0.10/0.22/0.43} 
& -0.14/0.02/0.13 \\
FastBEV++ 
& 0.04/0.05/0.11 
& 0.04/0.04/0.10 
& \textbf{0.11/0.29/0.47} \\
\bottomrule
\end{tabular}
\end{minipage}
\hfill
\begin{minipage}[t]{0.38\textwidth}
    \centering
\centering
\small
\setlength{\tabcolsep}{1pt}
\caption{Cross-vehicle-type transferability; reported 3D bbox displacement  $d_1/d_2/d_3$ (m) in three frames.}
\label{tab:transfer-type}
\renewcommand{\arraystretch}{1.0}
\begin{tabular}{lcc}
\toprule
\multirow{2}{*}{\shortstack{Source\\Vehicle\\Type}}  & \multicolumn{2}{c}{Target Vehicle Type} \\
\cmidrule(lr){2-3}
 & SUV & Sedan \\
\midrule
SUV   & \textbf{0.14}/\textbf{0.31}/\textbf{0.54} & 0.07/0.22/0.34 \\
Sedan & 0.09/0.23/0.35 & \textbf{0.10}/\textbf{0.26}/\textbf{0.51} \\
\bottomrule
\end{tabular}
\end{minipage}
\end{tabular}
\vspace{-.15in}
\end{table*}

\textbf{Cross-model transferability is challenging.}
Tab.~\ref{tab:transfer-model} reports the cross-model transferability results. Camouflage optimized on one detector generally fails to preserve strong displacement on another detector, with the best transferred final-frame displacement reaching only $0.13$~m. This limited transferability likely stems from detector-specific image-to-BEV transformations, feature aggregation strategies, and 3D box regression mechanisms, highlighting the fundamental challenge of inducing subtle box displacements under black-box settings.

\textbf{Cross-vehicle-type transfer remains effective.}
We further evaluate transferability across target-vehicle types. For each detector, we optimize the camouflage on one primary vehicle-type case and directly evaluate it on another after adapting the texture to the target vehicle body size and attachable surface. Tab.~\ref{tab:transfer-type} shows that the transferred camouflage still induces non-trivial 3D bbox displacement in consecutive frames when attached to another vehicle in another scenario.



\mysubsection{Ablation Study}
\label{sec:ablation}

\begin{figure}[t]
    \centering

    \begin{minipage}[c]{0.40\textwidth}
        \centering

        \begin{subfigure}[t]{0.48\textwidth}
            \centering
            \includegraphics[width=\textwidth,trim=600 0 0 0,,clip]{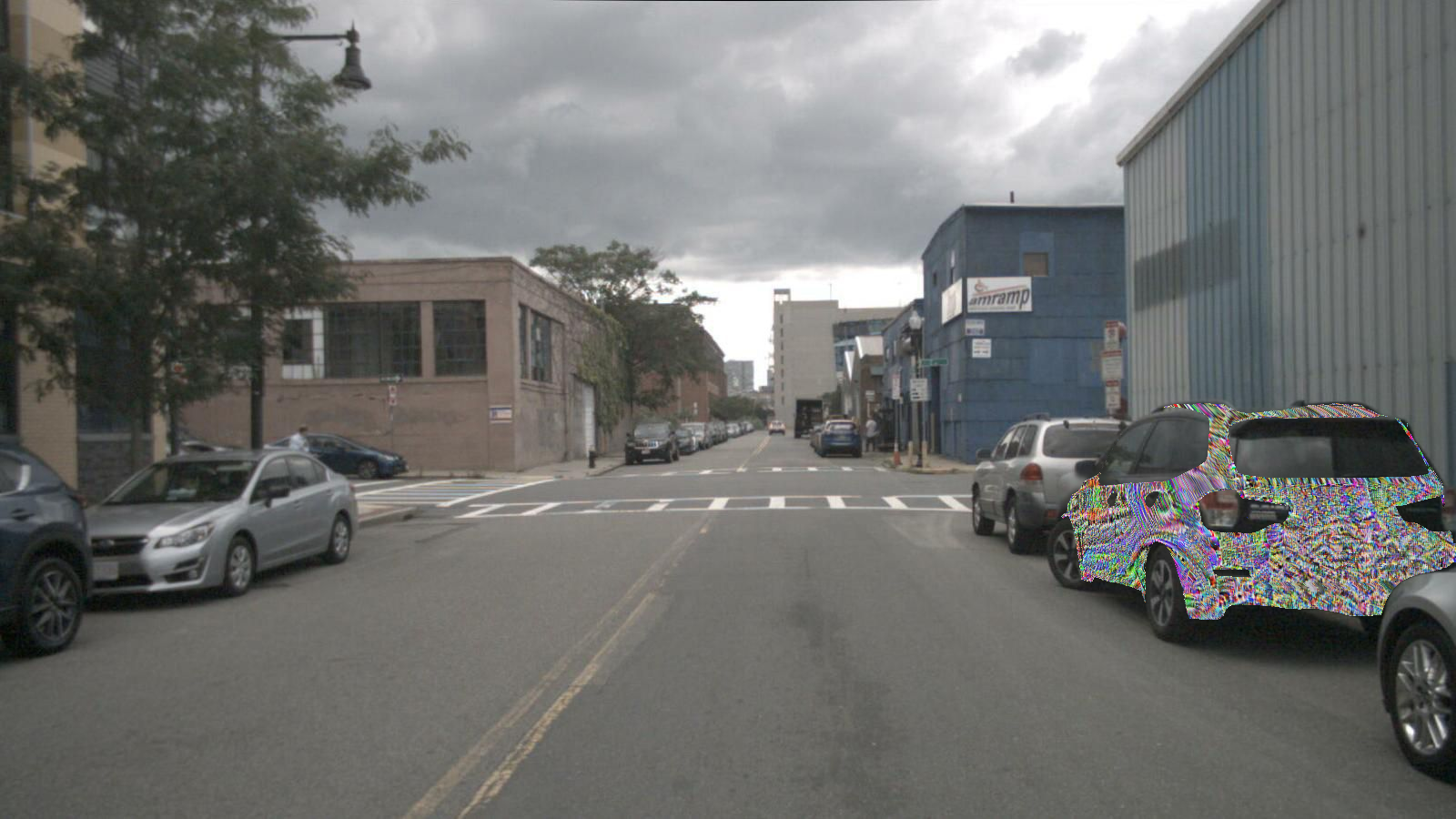}
            \caption{w/ $\mathcal{L}_{\mathrm{style}}$}
            \label{fig:wstyle}
        \end{subfigure}
        \hfill
        \begin{subfigure}[t]{0.48\textwidth}
            \centering
            \includegraphics[width=\textwidth,trim=600 0 0 0,,clip]{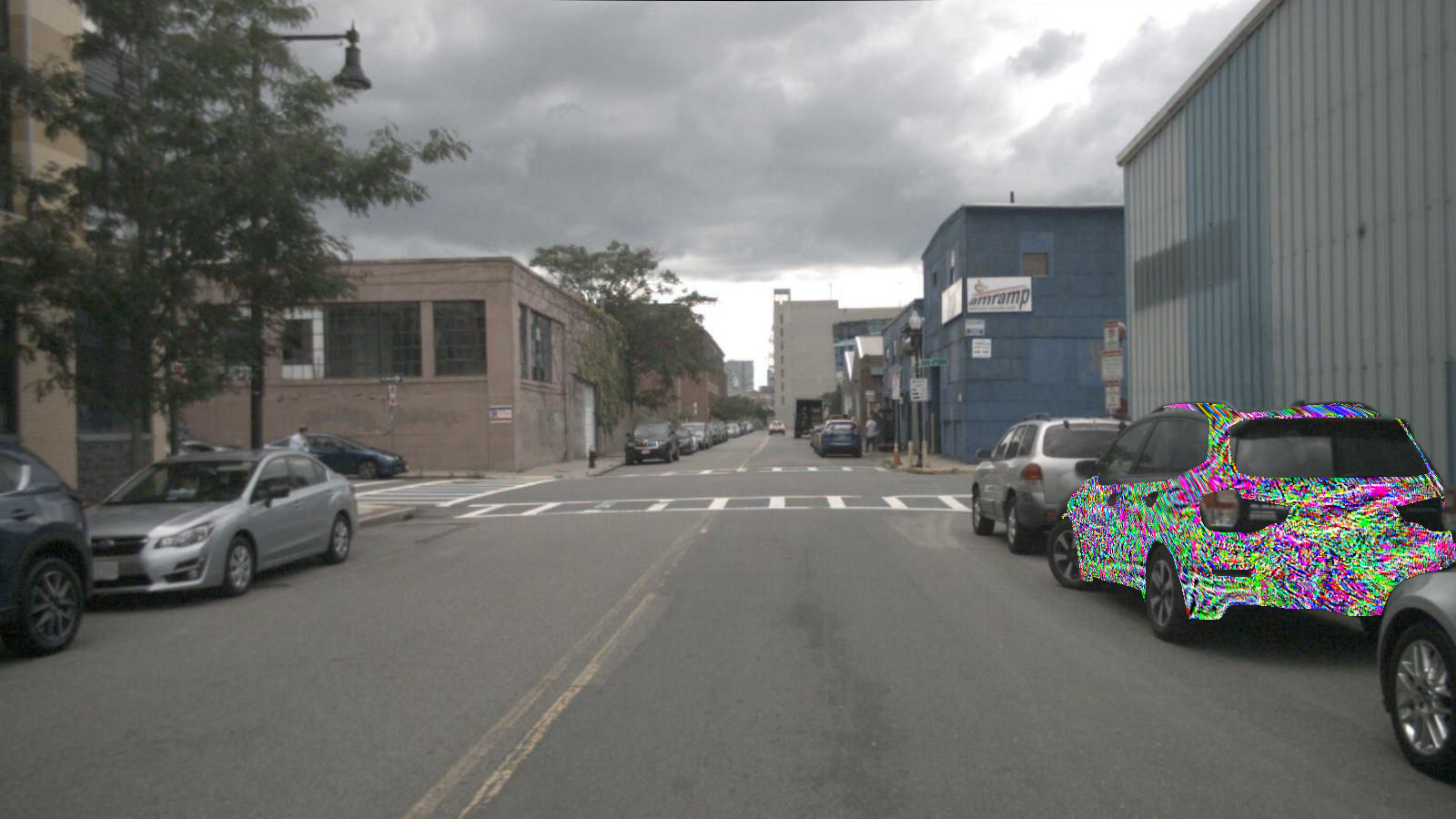}
            \caption{w/o $\mathcal{L}_{\mathrm{style}}$}
            \label{fig:wostyle}
        \end{subfigure}

        \caption{Ablation study of $\mathcal{L}_{\mathrm{style}}$.}
        \label{fig:style}
    \end{minipage}
    \hfill
    \begin{minipage}[c]{0.55\textwidth}
        \centering
        \small
        \setlength{\tabcolsep}{1pt}
        \captionof{table}{Ablation study of key attack components. ASR is reported in the format Pipeline A/B.}
        \label{tab:ablation}
        \renewcommand{\arraystretch}{1.0}
        \begin{tabular}{lccccc}
            \toprule
            Module & $d_3$ (m) $\uparrow$ & CV $\downarrow$ & PDR (\%) $\uparrow$ & BFS $\uparrow$ & ASR (\%) $\uparrow$ \\
            \midrule
            Full                    & \textbf{0.68} & 0.54          & \textbf{76.7} & 0.82          & \textbf{57.5/65.0} \\
            w/o VAF                 & 0.48          & 0.56          & 50.8          & 0.82          & 35.0/51.7 \\
            w/o EoT                 & 0.43          & 0.77          & 54.2          & 0.76          & 46.7/48.3 \\
            w/o $\mathcal{L}_{\mathrm{move}}$   & 0.12          & 1.05          & 51.7          & \textbf{0.93} & 15.8/23.3 \\
            w/o $\mathcal{L}_{\mathrm{prog}}$   & 0.51          & \textbf{0.48} & 58.3          & 0.83          & 26.7/41.7 \\
            w/o $\mathcal{L}_{\mathrm{fid}}$    & 0.64          & 2.04          & 65.0          & 0.60          & 49.2/57.5 \\
            \bottomrule
        \end{tabular}
    \end{minipage}
    \vspace{-.15in}
\end{figure}

Tab.~\ref{tab:ablation} reports the ablation results on the two primary cases using BEVDet as the victim detector. We use Coefficient of Variation (CV) to measure the stability of displacement across perturbation conditions, and Box Fidelity Score (BFS) to measure the similarity between attacked and clean boxes in target vehicle confidence and geometric attributes. BFS ranges from 0 to 1, where 1 indicates identical box fidelity. We draw the following conclusions:
(1) \textbf{Movement and progression objectives are both essential.}
The ablation results indicate that $\mathcal{L}_{\mathrm{move}}$ and $\mathcal{L}_{\mathrm{prog}}$ are the core attack objectives, respectively driving large bbox displacement and progressive temporal movement. 
(2) \textbf{VAF and EoT improve robustness and stability.}
Removing VAF degrades the overall performance, validating the importance of our viewing-angle-based design. EoT enhances attack robustness by reducing CV and improving the overall metrics. 
(3) \textbf{Fidelity constraints stabilize the attack.}
Removing $\mathcal{L}_{\mathrm{fid}}$ sharply reduces BFS and also lowers ASR, indicating that fidelity constraints help stabilize the optimization and preserve attack effectiveness.
The ablation effect of $\mathcal{L}_{\mathrm{style}}$ is shown in Fig.~\ref{fig:style}.

\mysubsection{Case Study}
\label{sec:case-study}

\begin{figure}[t]
    \centering



        \centering
        \includegraphics[width=\textwidth]{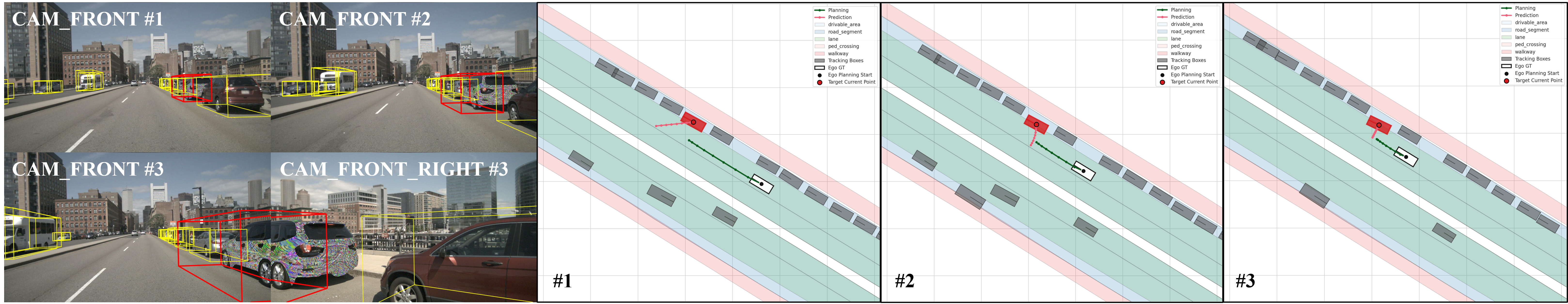}

    \caption{
    Representative case study of the attack results.
    The static camouflage induces progressive 3D bbox displacement and shifts the predicted target trajectory toward the ego future path, eventually triggering unnecessary braking.
    Left: camera-view observations, including three consecutive frames; Right: the resulted BEV prediction and planning results.}
    \vspace{-.15in}
    \label{fig:case}
\end{figure}

We present a representative case study of the end-to-end attack in Fig.~\ref{fig:case}. The static camouflage exhibits view-dependent appearance changes under relative ego-target motion, causing progressive 3D bbox displacement of $d1/d2/d3 = 0.31/0.63/0.91$~m, making the target appear to move toward the ego driving region, resulting in an APE of $2.82$~m and an MTD of $0.18$~m, and ultimately triggering hard braking by the ego vehicle. This confirms that the proposed attack is not a single-frame perception perturbation; instead, it exploits viewing-angle variation to turn static camouflage into temporally coherent 3D bbox shifts that propagate through downstream pipeline, causing unnecessary braking and supporting our motivation.
The abandoned overtaking scenario is provided in Appendix~\ref{sec:additonal-case}.

\section{Discussion}
\label{sec:discussion}

\textbf{Threats to validity.}
Our evaluation focuses on a limited set of perception models, tracking pipelines, and autonomous driving systems. Although the attack transfers across different settings, the effectiveness may vary for other architectures, sensor configurations, fusion strategies, and planning modules. In addition, our experiments are conducted on existing datasets and simulation-driven pipelines, which may not fully capture real-world deployment conditions.

\textbf{Mitigation and ethical consideration.}
Potential defenses include multi-sensor fusion~\cite{liu2023bevfusion,bai2022transfusion,li2022deepfusion,chen2023futr3d}, temporal and physical consistency reasoning~\cite{man2023person,physense}, and adversarial patch or camouflage defenses~\cite{chiang2021adversarial,liu2022segment,kim2022defending,jing2024pad}. However, prior work shows that fusion systems remain vulnerable to sensor attacks~\cite{fusionnoenough,cao2021invisible}, and these approaches mitigate but do not fundamentally eliminate the vulnerability exploited in this work: viewing-angle variation induces spatio-temporally coherent perception bias.
This work motivates the design of secure and robust autonomous driving systems by identifying this overlooked attack surface. We evaluate attacks on public datasets without harming real systems.

\section{Conclusion}

We present a view-induced trajectory manipulation attack that exploits natural viewing-angle variation to transform a static adversarial camouflage into temporally coherent motion deception. By inducing progressive 3D bounding-box displacement across consecutive frames, the attack propagates through downstream tracking, prediction, and planning modules, leading to harmful driving behaviors such as abandoned overtaking and unnecessary hard braking. Our results highlight viewing-angle variation as a new attack surface for autonomous driving systems and demonstrate the importance of evaluating security beyond single-frame perception robustness.

\bibliographystyle{plain}
\bibliography{reference}

\newpage
\appendix
\mysection{Additional Case Study}
\label{sec:additonal-case}

\begin{figure}[h]
    \centering
    \includegraphics[width=\textwidth]{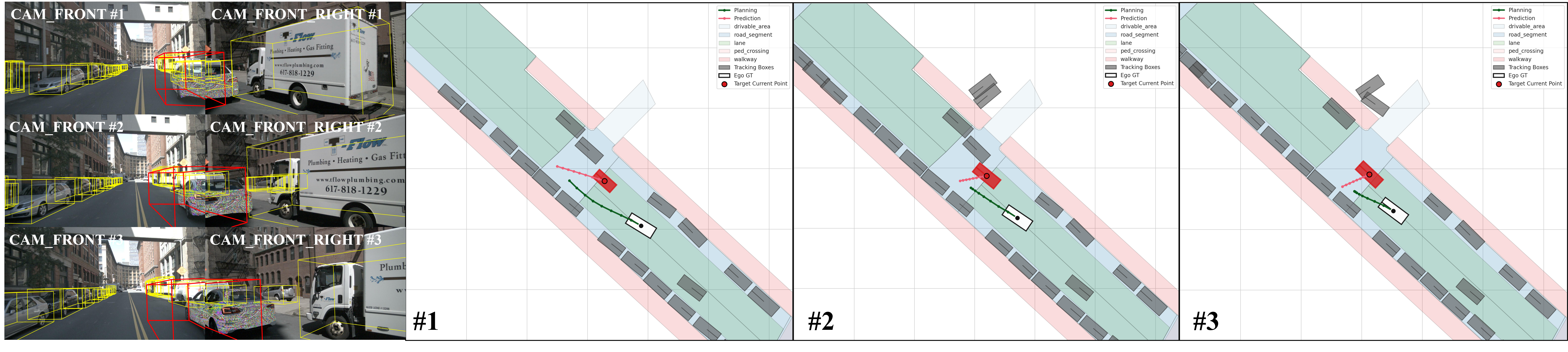}
    \caption{
    Additional case study of abandoned overtaking. In the clean scene, the target vehicle is preparing to turn right, and the ego vehicle plans to overtake it. Under attack, the induced 3D bbox displacement makes the victim AV perceive the target vehicle as drifting left toward the ego driving region, reducing the clearance required for overtaking and causing the planner to abandon the original overtaking maneuver. Left: camera-view observations, including three consecutive frames; Right: the resulted BEV prediction and planning results.}
    \vspace{-.15in}
    \label{fig:case-overtaking}
\end{figure}

In addition to the hard-braking case in the main text, we provide an additional case study of abandoned overtaking in Fig.~\ref{fig:case-overtaking}. In the clean scene, the ego vehicle is planning to overtake the target vehicle and continue along the original driving region. Under attack, the induced 3D bbox displacement increases across frames with $d_1/d_2/d_3=\text{0.42}/\text{0.77}/\text{1.16}$~m, shifting the predicted target trajectory closer to the ego driving region and reducing the MTD to $0.05$~m, with an APE of $2.80$~m. Consequently, the victim AV abandons the overtaking maneuver.

\end{document}